\documentclass[12pt]{article}
\usepackage{amsmath,amssymb}
\usepackage{latexsym,graphicx,verbatim}

\usepackage[usenames]{color}

\def\6#1{{\underline{#1}}}
\def\m6#1{{\underline{#1}\,}}

\newdimen\Tdim
\def\ispan{{\setbox0=\hbox{i}%
\Tdim\ht0\advance\Tdim\dp0\rule[-\dp0]{0pt}{\Tdim}}}
\def\jspan{{\setbox0=\hbox{j}%
\Tdim\ht0\advance\Tdim\dp0\rule[-\dp0]{0pt}{\Tdim}}}
\def\Tspan#1{{\setbox0=\hbox{#1}%
\Tdim\ht0\advance\Tdim\dp0\advance\Tdim.55ex\rule[-\dp0]{0pt}{\Tdim}\box0}}

\def\be{\begin{eqnarray}}
\def\ben{\begin{eqnarray*}}
\def\ee{\end{eqnarray}}
\def\een{\end{eqnarray*}}
\def\Tr{{\rm Tr}}

\def\p{\partial}

\def\=:{=\hspace{-.7em}\raisebox{1.1ex}{.}\hspace{.1em}\raisebox{-0.2ex}{.} }

\newcommand {\beq}{\begin{eqnarray}}
\newcommand {\eeq}{\end{eqnarray}}
\newcommand {\non}{\nonumber\\}

\def\mn2changed#1{\textcolor{blue}{{\bf #1}}}

\makeatletter
\@addtoreset{equation}{section}

\renewcommand{\thefootnote}{\fnsymbol{footnote}}
\makeatother

\makeatletter
\newcommand{\thetablename}{Table}
\def\fnum@table{\thetablename\ \thetable}
\makeatother

\begin{document}
\thispagestyle{empty}
\begin{flushright}
IFUP-TH/2009-6\\
{\tt arXiv:0903.1528} \\
March, 2009 \\
\end{flushright}
\vspace{3mm}
\begin{center}
{\LARGE  
Non-Abelian Global Vortices
} \\ 
\vspace{20mm}

{\normalsize
Minoru~Eto$^{a,b}$, 
Eiji~Nakano$^c$ and
Muneto~Nitta$^d$}
\footnotetext{
e-mail~addresses: \tt
minoru(at)df.unipi.it;
e.nakano(at)gsi.de;
nitta(at)phys-h.keio.ac.jp.
}

\vskip 1.5em
{\footnotesize
$^a$ {\it INFN, Sezione di Pisa,
Largo Pontecorvo, 3, Ed. C, 56127 Pisa, Italy
}
\\
$^b$ {\it Department of Physics, University of Pisa
Largo Pontecorvo, 3,   Ed. C,  56127 Pisa, Italy
}
\\
$^c$ {\it 
Extreme Matter Institute, GSI, Planckstr. 1, D-64291 Darmstadt, Germany
}
\\
$^d$ 
{\it Department of Physics, Keio University, Hiyoshi, Yokohama,
Kanagawa 223-8521, Japan
}
}
 \vspace{12mm}

%%%%%%%%%%%%%%%%%%%%%%%%%%%%%%%%%%%%%%%%%%%%%%%%%%
\abstract{
We study topologically stable 
non-Abelian global vortices in the $U(N)$ linear sigma model. 
The profile functions of the solutions are numerically obtained. 
We investigate the behaviour of vortices in 
two limits in which masses of traceless or trace 
parts of massive bosons are much larger than the others. 
In the limit that the traceless parts are much heavier, 
we find a somewhat bizarre vortex solution 
carrying a non-integer $U(1)$ winding number $1/\sqrt N$  
which is {\it irrational} in general.
}
%%%%%%%%%%%%%%%%%%%%%%%%%%%%%%%%%%%%%%%%%%%%%%%%%%

\end{center}

\vfill
\newpage
\setcounter{page}{1}
\setcounter{footnote}{0}
\renewcommand{\thefootnote}{\arabic{footnote}}

%%%%%%%%%%%%%%%%%%%%%%%%%%%%%%%%%%%%%%%%%%%%%%%%%%%%%%%%%%%%%%%%%%%%%%%%%%%%%%%%
%%%%%%%%%%%%%%%%%%%%%%%%%%%%%%%%%%%%%%%%%%%%%%%%%%%%%%%%%%%%%%%%%%%%%%%%%%%%%%%%

\section{Introduction}
Superfluid vortices appear in various condensed matter systems 
such as helium superfluid. 
They are global vortices in relativistic field theories \cite{Davis:1989gn}.
When a global $U(1)$ symmetry is spontaneously broken, 
for instance by the order parameter $\left<\phi\right>=v$ 
with $\phi$ a complex scalar field in the Goldstone model,
a $U(1)$ Nambu-Goldstone boson appears. 
Then in general there appears a global string 
asymptotically $\phi \sim v e^{i \theta}$ 
winding around the vacuum manifold (or the order parameter space) $U(1)$.
The energy of global strings is 
logarithmically divergent in the infinitely large space, 
unlike local vortices with gauged $U(1)$.
So less attention have been paid to global vortices 
as a candidate of cosmic strings 
compared with local counter part. 
When the net number of strings is zero, for instance, 
a pair of string and anti-string or string loops, their energy is finite and can be considered 
as cosmic strings \cite{cosmic,Vilenkin}. 
Axion strings are such objects. 

Natural non-Abelian extension of $U(1)$ vortices is 
$U(N)$ vortices in the $U(N)$ linear sigma model 
for which the order parameter is extended to 
an $N$ by $N$ complex matrix $\left<\Phi\right> = v {\bf 1}_N$.\footnote{
Other non-Abelian global vortices appear in 
the B-phase of $^3$He in which symmetry is broken as 
$SO(3)_{\rm S} \times SO(3)_{\rm L} \times U(1) \to SO(3)_{\rm S+L}$. 
In this case the corresponding vacuum manifold is $U(1) \times SO(3)$ 
and the first homotopy group is 
 $\pi_1[U(1) \times SO(3)] \simeq {\mathbb Z} \oplus {\mathbb Z}_2$.
} 
Such non-Abelian vortices are expected to form 
during the chiral phase transition in QCD, 
in which case the field $\Phi$ 
of a 3 by 3 matrix ($N=3$)
is a condensate of quark-anti-quark 
$\left<\Phi\right> \sim \left<\bar q q\right>$.
At the chiral phase transition \cite{Pisarski:1983ms},  
the chiral symmetry $SU(N)_{\rm L} \times SU(N)_{\rm R}$ 
is spontaneously broken down to its diagonal symmetry 
$SU(N)_{\rm V}$. 
According to this breaking, massless Nambu-Goldstone bosons 
appear as pions (or more generally mesons).
At the same time, the axial symmetry 
$U(1)_{\rm A}$ is also spontaneously broken 
and the $\eta'$ meson appears. 
However
$U(1)_{\rm A}$ is explicitly broken by the axial anomaly at 
zero temperature, 
giving a mass to the $\eta'$ meson.
It has been argued that the axial anomaly might disappear 
and $U(1)_{\rm A}$ is approximately 
recovered at high temperature \cite{anomaly}. 
Although there is still an ambiguity if it occurs below 
the temperature of the chiral phase transition, 
let us consider such a situation. 
Then the breaking pattern is 
$U(1)_{\rm A} \times SU(N)_{\rm L} \times SU(N)_{\rm R} 
\to SU(N)_{\rm V}$ apart from the discrete symmetry, 
and the vacuum manifold is 
\beq
{SU(N)_{\rm A} \times U(1)_{\rm A} \over {\mathbb Z}_N}
\simeq U(N)_{\rm A} .\label{eq:vac_mfd}
\eeq
The first homotopy group $\pi_1[U(N)_{\rm A}] \simeq {\mathbb Z}$
is non-trivial and so there exist topologically stable 
vortex-strings in this breaking. 
The simplest vortex appearing  is 
a $U(1)$ vortex-string called the $\eta'$ string 
asymptotically given by 
$\Phi \sim v e^{i\theta} {\bf 1}_N$, 
which winds around $U(1)_{\rm A}$ once ($2\pi$)
\cite{Zhang:1997is,Brandenberger:1998ew}.\footnote{
Brandenberger {\it et.~al} have discussed non-topological strings, 
called the pion strings, 
using massless Nambu-Goldstone (pions) $SU(2)_{\rm A}$ 
\cite{Zhang:1997is,Brandenberger:1998ew,Nagasawa:1999iv}.
They are topologically unstable 
because of $\pi_1[SU(N)]=0$. 
In this sense non-Abelian strings below are made of 
both $\eta'$ mesons and pions.
}
However this is not the minimum vortex-string because 
there exists a smaller loop in the vacuum manifold (\ref{eq:vac_mfd}); 
The minimum string is a non-Abelian string asymptotically given by
$\Phi \sim v {\rm diag}\ (e^{i\theta},1,\cdots,1)=
v\,e^{i\frac{\theta}{N}} 
{\rm diag}\left(e^{i\frac{N-1}{N}\theta},e^{-i\frac{\theta}{N}},
\cdots,e^{-i\frac{\theta}{N}}\right)$, which winds $U(1)_{\rm A}$ as well as 
the $SU(N)_{\rm A}$ \cite{Balachandran:2002je}.
The important is that this string 
winds $2\pi/N$ of $U(1)_{\rm A}$ and 
therefore its tension is $1/N$ of 
that of the $U(1)$ $\eta'$-string.
Such vortices with fractional $U(1)$ winding number often 
appear in various condensed matter systems 
such as Bose-Einstein condensates and 
certain types of superconductors, 
and are called ``fractional vortices".\footnote{
For instance in the polar phase 
of a spin 1 spinor Bose-Einstein condensate, 
$U(1)_{\Phi} \times SO(3)_{S}$ is spontaneously 
broken down to 
$U(1)_{\Phi+S} \rtimes ({\mathbb Z}_2)_{\Phi+S} $ 
with $\rtimes$ denoting a semi-direct product. 
Then the vacuum manifold is $M \simeq [U(1)_{\rm \Phi} \times SO(3)_{\rm S}]
/[U(1)_{\Phi+S} \rtimes ({\mathbb Z}_2)_{\Phi+S}] \simeq [U(1) \times S^2]/{\mathbb Z}_2$ 
\cite{spin1-OPS} 
and the first homotopy group $\pi_1(M) \simeq {\mathbb Z}$ 
supports half quantized vortices \cite{spin1-vortex}.
Similarly to this, 1/3 quantized vortices exist in the cyclic phase 
of a spin 2 spinor Bose-Einstein condensate \cite{spin2-vortex}.
}

At low temperature the axial anomaly  
induces the periodic potential in $U(1)_{\rm A}$.
The $U(1)_{\rm A}$ is broken to ${\mathbb Z}_N$ and
there appear $N$ disconnected vacua, 
in each of which the $\eta'$ meson gets mass.
A $U(1)$ $\eta'$-string is accompanied with 
$N$ domain walls and the 
total configuration becomes $N$ domain wall junction 
with a string at the junction line \cite{Balachandran:2001qn}. 
Balachandran {\it et.~al} have discussed 
a possible role of such an object 
in the early universe \cite{Balachandran:2001qn}. 

The presence of a non-Abelian vortex breaks 
the $SU(N)_{\rm V}$ symmetry of vacua to its subgroup 
$SU(N-1)_{\rm V} \times U(1)_{\rm V}$ and consequently 
there appear further Nambu-Goldstone modes 
${\mathbb C}P^{N-1}\simeq SU(N)/[SU(N-1) \times U(1)]$ 
which are orientations in the internal space \cite{Nitta:2007dp}.
This idea was brought from 
the local $U(N)$ vortices \cite{Hanany:2003hp}
for which $U(1)_{\rm A}$ and $SU(N)_{\rm L}$ are gauged.
However there exists a crucial difference between 
global and local $U(N)$ vortices.
The Nambu-Goldstone modes ${\mathbb C}P^{N-1}$ 
of local $U(N)$ vortices are localized around the vortex 
and become the moduli (or collective coordinates) of the vortex 
\cite{Hanany:2003hp}
while those of global $U(N)$ vortices are not 
localized but spread to infinity 
(or the boundary of a finite space). 
Having this in mind, 
an inter-string force between two parallel global $U(N)$ vortex-strings with 
different orientations in the internal space 
has been calculated recently \cite{Nakano:2007dq,Nakano:2008dc}. 
The force depends on the relative orientation: 
it reaches the maximum  
when two strings wind around the same component of the 
vacuum expectation value  
[for instance ${\rm diag}\ (e^{i\theta},1,\cdots,1)$ 
and ${\rm diag}\ (e^{i\theta},1,\cdots,1)$], 
and it vanishes when the two strings wind around different components of the 
vacuum expectation value  
[for instance ${\rm diag}\ (e^{i\theta},1,1,\cdots,1)$ 
and ${\rm diag}\ (1,e^{i\theta},1,\cdots,1)$]. 
This result implies that 
a $U(1)$ $\eta'$-string is marginally decomposed into 
$N$ pieces of non-Abelian strings: 
$e^{i\theta}{\bf 1}_N \to 
{\rm diag}\ (e^{i\theta},1,1,\cdots,1) 
+ {\rm diag}\ (1,e^{i\theta},1,\cdots,1) + \cdots$.
Such decomposition necessary occurs at finite temperature 
where the free energy is minimized instead of the energy. 
Therefore at low temperature with the axial anomaly, 
a domain wall junction \cite{Balachandran:2001qn} is unstable, 
because a $U(1)$ string is pulled by each of $N$ domain walls
and is decomposed into $N$ non-Abelian strings 
to each of which one domain wall is attached. 
In the end each piece is pulled to infinity in each of $N$ directions.\footnote{Stable domain wall junctions or networks exist 
in gauged $U(N)$ linear sigma models with appropriate masses 
\cite{Eto:2005cp}.} 

In this paper we study the purely string solution without domain walls 
in the linear sigma model without the axial anomaly. 
Numerical solutions themselves were previously obtained 
in \cite{Nitta:2007dp}.
Here we study profile functions of solutions 
in much more detail with more accuracy. 
By using the relaxation method, we numerically determine 
the shooting parameters of the solutions up to fifth order. 
We investigate the dependence of the profiles of the vortex 
to the parameters in the linear sigma model.
We also study the two limits in which 
the masses of traceless or trace parts of the massive bosons 
in the linear sigma model 
are much larger than the others. 
In these limits, the model reduces some nonlinear sigma models. 
In the limit that the trace parts are much heavier, 
the equations for the profiles of the $U(N)$ vortex 
become sine-Gordon-like.
The solution remains to be regular. 
On the other hand, in the limit that the traceless parts are much heavier, 
we find somewhat surprising solution; 
the $U(N)$ vortex solution 
reduces to a singular $U(1)$ vortex 
with the $U(1)$ winding number $1/\sqrt N$ 
which is {\it irrational} 
specifically for $N=2,3,5,6, \cdots$.
In general, 
profiles of $U(1)$ vortices with a non-integer $U(1)$ winding number ($\le 1$) 
are of course singular.
Interesting is that 
such an ``{\it irrational vortex}" naturally appears 
in a particular limit of a regular non-Abelian vortex solution.
As far as we know such a vortex has not been reported yet in the literature. 

This paper is organized as follows. 
In Sec.~\ref{sec:U(1)} we review the $U(1)$ global vortex solution 
in the Goldstone model. We give numerical solutions and 
study their asymptotic behaviours. 
In Sec.~\ref{sec:U(N)} we study 
the non-Abelian vortex solution in the $U(N)$ linear sigma model.
We derive profile functions of the minimum $U(N)$ vortex 
with the $U(1)$ winding number $1/N$ for 
$N=2,3,\cdots,10$ 
and determine the shooting parameters up to fifth order. 
We also discuss various limits by sending some of 
masses to infinity. 
We find that in a particular limit the non-Abelian $U(N)$ vortex 
reduces to an Abelian vortex with the irrational $U(1)$ winding number 
$1/\sqrt N$. 
Sec.~\ref{sec:conc} is devoted to conclusion and discussion.
Behaviors of vortex solutions in the large $N$ limit is discussed 
in Appendix.

%\newpage
%%%%%%%%%%%%%%%%%%%%%%%%%%%%%%%%%%%%%%%%%%
\section{Global $U(1)$ Vortices}\label{sec:U(1)}

\subsection{The Goldstone Model and Vortex Solutions}

Let us begin with giving a review on a global vortex-string solution in the Goldstone model with 
a complex scalar field $\phi(x)$
\beq
{\cal L} = |\p_\mu \phi|^2 - \lambda \left(|\phi|^2 - v^2\right)^2.
\label{eq:lag_goldstone}
\eeq
We choose $\lambda > 0$ and $v^2 >0$ for stable vacua with 
broken $U(1)$ symmetry.
The scalar potential is like a wine bottle, so we have an $S^1$ vacuum space with radius $|\phi| = v$.
When we choose a vacuum $\phi = v$ and consider small fluctuations as $\phi = v + \frac{\varphi_1 + i\varphi_2}{\sqrt 2}$
($\varphi_{1,2} \in \mathbb{R}$),
the Lagrangian in terms of the small fluctuations 
up to quadratic terms is of the form
\beq
{\cal L}^{(2)} = \frac{1}{2}(\p_\mu\varphi_1)^2 
+ \frac{1}{2}(\p_\mu\varphi_2)^2 
- 2\lambda v^2 \varphi_1^2.
\eeq
This shows that $\varphi_1$ is massive and $\varphi_2$ is a massless Nambu-Goldstone mode: 
\beq
m_1^2 \equiv 4 \lambda v^2,\quad m_2^2 = 0.
\eeq

\begin{figure}[ht]
\begin{center}
\begin{tabular}{cc}
\includegraphics[width=8cm]{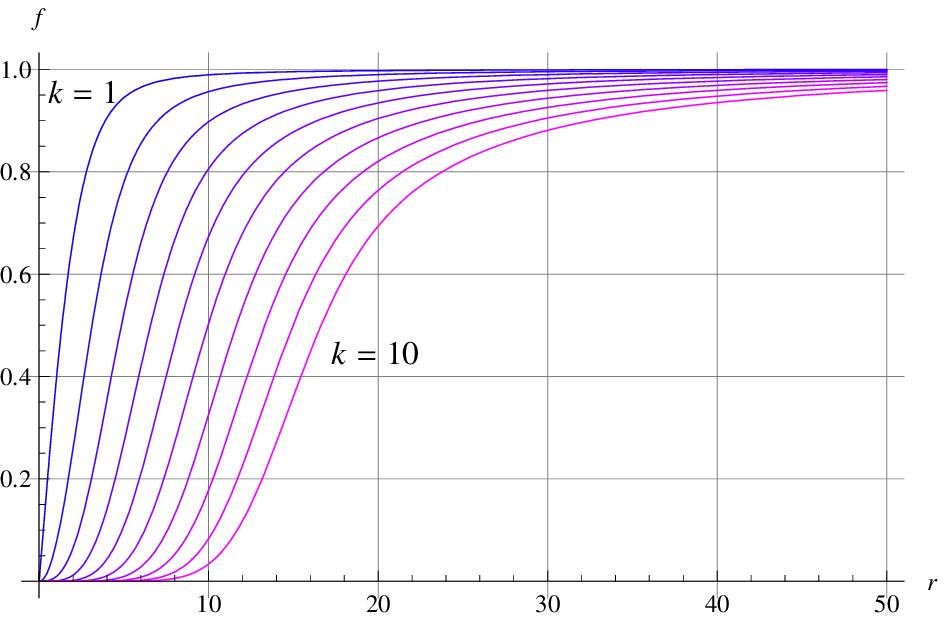} & 
\includegraphics[width=8cm]{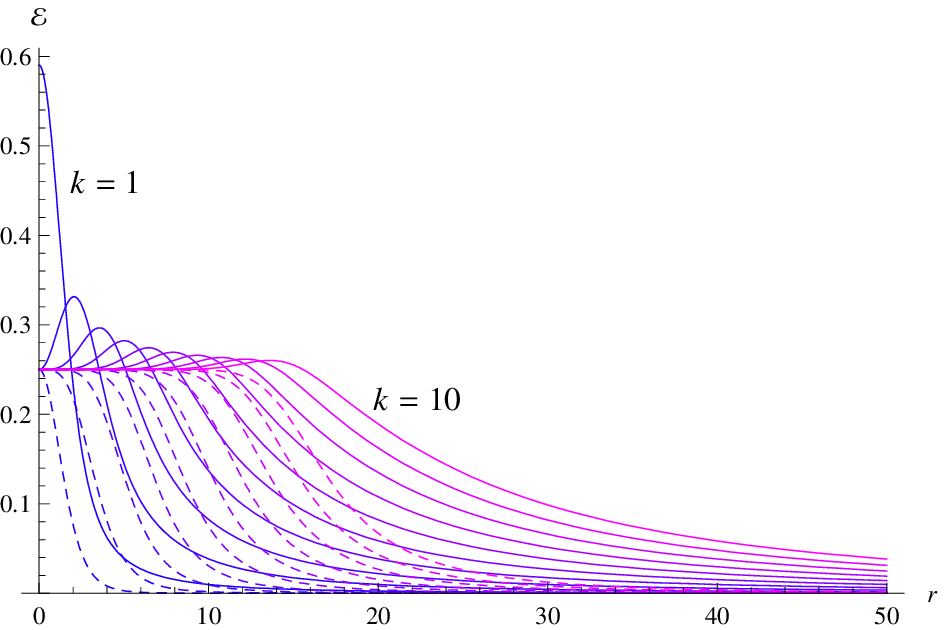}
\end{tabular}
\caption{Left panel shows the profile functions and the corresponding energy
densities ${\cal E}(k)$ (solid lines) and ${\cal E}_{\rm pot}(k)$ (broken lines) are plotted in the right panel for
$k=1,2,\cdots, 10$ and $m_1=1$.}
\label{fig:gv0}
\end{center}
\end{figure}

The equation of motion of $\phi$ reads
\beq
\p_\mu\p^\mu \phi + 2 \lambda \phi (|\phi|^2 - v^2) = 0.
\label{eq:eom_gv0}
\eeq
A global vortex-string extending linearly to the $x_3$-axis is obtained 
by solving (\ref{eq:eom_gv0}) with
an axisymmetric vortex ansatz
in the cylindrical coordinates $x_1 + i x_2 = r e^{i\theta}$, given by
\beq
\phi(r,\theta) = v\,e^{ik\theta} f(r),\quad k \in \mathbb Z,
\label{eq:ansatz_gv}
\eeq
with the boundary conditions
\beq
\lim_{r \to \infty} f(r) = 1,\quad
\lim_{r \to 0} f(r) = 0.
\label{eq:bc}
\eeq 
Plugging the ansatz (\ref{eq:ansatz_gv}) 
into Eq.~(\ref{eq:eom_gv0}), we get a second order differential equation
\beq
f'' + \frac{f'}{r} - \frac{k^2 f}{r^2} 
- \frac{m_1^2}{2}  f\left(f^2 - 1\right) = 0.
\label{eq:agv}
\eeq
Numerical solutions for $k=1,2,\cdots,10$ 
with the boundary conditions (\ref{eq:bc})
are plotted 
in the left panel of Fig.~\ref{fig:gv0}.

The energy of the vortex solution can be expressed as
\beq
E = 2\pi \int_0^\infty dr\, r {\cal E} 
= 2\pi v^2 \int_0^\infty dr\, r
\left[
f'{}^2 +\frac{k^2 f^2}{r^2}+\frac{m_1^2}{4}(f^2 -1 )^2
\right],
\label{eq:energy_u1}
\eeq
where the first two terms come from the derivative of the field $\phi$ and
the last term is from the scalar potential. Since $f \to 1$ as $r\to\infty$,
the kinetic energy logarithmically diverges. So the energy of the vortex-string consists of a finite
part and a logarithmically divergent part as
\beq
E(k) =  E_{\rm der}(k) + E_{\rm div}(k) + E_{\rm pot}(k)
\eeq
where we have defined
\beq
E_{\rm der}(k) &\equiv& 2\pi v^2 \int_0^\infty dr\, r
f'{}^2 ,\\
E_{\rm pot}(k) &=& 2\pi \int_0^\infty dr\, r {\cal E}_{\rm pot}
  \equiv \frac{\pi v^2m_1^2}{2} \int_0^\infty dr\, r (f^2 -1 )^2,\\
E_{\rm div}(k) &\equiv& 2 \pi v^2 k^2 \int^\infty_0 dr\,\frac{f^2}{r} = {\rm const.} 
+ 2\pi v^2 k^2 \lim_{L\to\infty} \log \frac{L}{r_0},
\eeq
Here we have introduced an IR cut-off $L$ 
which is the size of the system
and $r_0 \simeq m_1^{-1}$ is a typical size of the $U(1)$ global vortex such as
$|f - 1| \ll 1$ as $r \gg r_0$. 
We can analytically calculate $E_{\rm pot}$ as follows.
Let us first introduce 
a dimensionless coordinate $\rho \equiv m_1 r$.
The equation becomes independent of the coupling constants
\beq
F[f(\rho);k] \equiv 
f'' + \frac{f'}{\rho} - \frac{k^2 f}{\rho^2} 
- \frac{1}{2}  f \left(f^2 - 1\right) = 0.
\label{eq:eom_gv}
\eeq
Then we see that $E_{\rm pot}$ is independent of the scalar mass $m_1$:
\beq
E_{\rm pot}(k) = \frac{\pi v^2 m_1^2}{2} \int_0^\infty dr\, r (f^2-1)^2 = 
2\pi v^2 \int_0^\infty d\rho\, \rho \frac{(f^2-1)^2}{4}.
\eeq
To calculate this, let us make a trick
\beq
0 = 2 \rho^2 f'F[f;k] = \frac{d}{d\rho}\left[
\rho^2 f'{}^2 - k^2 f^2 - \frac{\rho^2}{4}(f^2-1)^2 
\right] + \frac{\rho}{2}(f^2 - 1)^2.
\eeq
By using this equation, we can bring $E_{\rm pot}$ in the following form
\beq
E_{\rm pot}(k) = \pi v^2 \left[
\rho^2 \left( - f'{}^2 + \frac{1}{4}(f^2-1)^2 \right) + k^2 f^2 
\right]^\infty_0 = \pi v^2 k^2.
\eeq
This formula is called the Derrick-Pohozaev identity 
in the literature \cite{Manton:2004tk}.
To derive the rightmost equality, 
we have used the asymptotic behaviours $f \propto \rho^k$ at $\rho \ll 1$ 
and $f = 1 - a_2/\rho^2 + {\cal O}(\rho^{-4})$ for $\rho \gg 1$, 
which will be obtained in the next subsection. 
We plot the total and potential energy densities ${\cal E}$ and 
 ${\cal E}_{\rm pot}$, respectively  
of our numerical solutions in the right panel of Fig.~\ref{fig:gv0}.
We also have numerically checked 
$E_{\rm pot}/(\pi v^2 k^2) = 0.999 + {\cal O}(10^{-4})$.
We can show that $E_{\rm der}$ is also finite. 

Note that higher winding solutions ($k\ge2$), especially co-axial vortices as we assumed above, are
not stable. 
Since the energy of the co-axial vortices is almost proportional to $k^2$, distant vortices 
are energetically preferred ($k^2 > k$ for $k \ge 2$). 
Therefore, the above static solutions with $k > 1$ are artifacts of
our co-axial ansatz and boundary conditions. 

%%%%%%%%%%%%%%%%%%%%%%%%%%%%%%%%%%%%%%%%%%%%%%%%%%%%%%
\subsection{Asymptotic behaviors at $r \to 0$ and $r\to \infty$}

Let us investigate asymptotic behavior at $m_1 r = \rho  \ll 1$ where 
the profile function is very small $|f|  \ll 1$. We expand the profile function as
\beq
f(\rho) = \sum_{n=1}^\infty {a_n} \rho^n.
\label{eq:exp1}
\eeq
By substituting this in Eq.~(\ref{eq:eom_gv}), we find that
the first non-zero coefficient is that of $\rho^k$
\beq
a_k = \lim_{\rho \to 0}\frac{f(\rho)}{\rho^k}
\eeq
which is sometimes called the shooting parameter and 
which may be determined by making use of
numerical solutions. Such parameters are important since we can uniquely determine solutions
with the parameters.
In the minimal winding ($k=1$) vortex, we got $a_1 = 0.4123772$ and
\beq
f_{k=1}(\rho) = a_1 \rho -\frac{a_1}{16} \rho^3 +\frac{a_1+16 a_1^3}{768} \rho^5
- \frac{a_1+160 a_1^3}{73728} \rho^7 + {\cal O}(\rho^9).
\eeq
In principle, one can infinitely increase accuracy of the approximation with the unique parameter $a_1$.
Although we assumed $\rho \ll 1$ in the beginning, 
we can reach at $\rho_0  \gtrsim 1$ with a good accuracy 
by increasing the order of expansion.

Let us next discuss asymptotics at $r \to \infty$.
We expand solution in the following way
\beq
f(\rho) = 1 - \sum_{i=1}^{\infty} \frac{b_i}{\rho^i},\quad {\rm for}\quad \rho \gg 1,
\label{eq:exp2}
\eeq
$b_i$ being constant. Unlike the expansion parameters $\{a_i\}$ in Eq.~(\ref{eq:exp1}),
the expansion parameters $\{b_i\}$ can be precisely determined. To this end, we insert 
(\ref{eq:exp2}) into $F[f;k]$ given in Eq.~(\ref{eq:eom_gv}) and determine $b_i$ by comparing terms
order by order. Then we get $b_{\rm odd} = 0$ and
\beq
b_2 = k^2,\quad
b_4 = \frac{1}{2} k^2 \left(8+k^2\right),\quad
b_6 = \frac{1}{2} k^2 \left(128+32 k^2+k^4\right),\quad \cdots 
\eeq
We can go on up to order which we desire.

%%%%%%%%%%%%%%%%%%%%%%%%%%%%%%%%%%%%%%%%%%%%%%%%%%%%
\section{Global $U(N)$ Vortices} \label{sec:U(N)}

Let us study non-Abelian global vortices.
Basic analysis on them have been done in \cite{Nitta:2007dp}. 
Here we are going to push forward
the analysis of \cite{Nitta:2007dp} on vortex solutions 
by investigating the equations of motion in much detail.

%%%%%%%%%%%%%%%%%%%%%%%%%%%%%%%%%%%%%%%%%%%%%%%%%%%
\subsection{The $U(N)$ Linear Sigma Model and Vortex Solutions}

The model that we consider here is a natural extension of 
the Goldstone model given in Eq.~(\ref{eq:lag_goldstone}). 
It is the 
$SU(N)_{\rm L} \times SU(N)_{\rm R} \times U(1)_{\rm A}$ linear sigma model
for an $N \times N$ complex matrix of scalar fields $\Phi(x)$, given by   
\beq
{\cal L} = \Tr\left[\p_\mu\Phi^\dagger\p^\mu\Phi 
- \lambda_2 (\Phi^\dagger\Phi)^2 + \mu^2 \Phi^\dagger\Phi 
\right]
- \lambda_1\left(\Tr[\Phi^\dagger\Phi]\right)^2 - \frac{\mu^4 N}{4(N\lambda_1 + \lambda_2)} 
\label{eq:lsm}
\eeq
where the last constant term is introduced for 
the vacuum energy to vanish. 
For a stability of vacua, we consider the parameter region 
$\mu^2 > 0$, $\lambda_2 > 0$ and $N\lambda_1 + \lambda_2 > 0$. 
The chiral symmetry $SU(N)_{\rm L} \times SU(N)_{\rm R}$ 
and the axial symmetry $U(1)_{\rm A}$ act on $\Phi$ as
\beq
 \Phi \to e^{i\theta} g_L \Phi g_R  , \quad
 (e^{i\theta}, g_L, g_R ) 
 \in U(1)_{\rm A} \times SU(N)_{\rm L} \times SU(N)_{\rm R}. 
 \label{eq:chiral}
\eeq
However, unlike the usual case in the absence of 
$U(1)_{\rm A}$ broken by the axial anomaly,  
the structure of discrete symmetries becomes somewhat complicated
in the presence of $U(1)_{\rm A}$. 
Here we explain it in detail.
First the group action $G$ on $\Phi$ is not Eq.~(\ref{eq:chiral}) itself 
but is given by
\beq
 G = { U(1)_{\rm A} \times SU(N)_{\rm L} \times SU(N)_{\rm R} 
      \over {\mathbb Z}_N \times {\mathbb Z}_N}, 
\eeq
where the following ${\mathbb Z}_N \times {\mathbb Z}_N$ action 
does not act on $\Phi$ and therefore is removed from $G$: 
\beq
  && (\omega^{-k-l},\omega^k {\bf 1}_N,\omega^l {\bf 1}_N) 
    \in U(1)_{\rm A} \times SU(N)_{\rm L} \times SU(N)_{\rm R},\nonumber\\
 && \omega \equiv e^{2\pi i/N},\quad (k,l=0,1,2,\cdots,N-1). 
  \label{eq:ZNxZN}
\eeq
For later use let us redefine these discrete groups as 
${\mathbb Z}_N \times {\mathbb Z}_N \simeq 
({\mathbb Z}_N)_{\rm V} \times ({\mathbb Z}_N)_{\rm A}$ with
\beq
&& ({\mathbb Z}_N)_{\rm V}: 
  (1,\omega^k {\bf 1}_N,\omega^{-k} {\bf 1}_N) 
    \in U(1)_{\rm A} \times SU(N)_{\rm L} \times SU(N)_{\rm R} 
  \label{eq:ZNV} \\
&& ({\mathbb Z}_N)_{\rm A}: 
  (\omega^{-2k},\omega^k {\bf 1}_N,\omega^k {\bf 1}_N) 
    \in U(1)_{\rm A} \times SU(N)_{\rm L} \times SU(N)_{\rm R}  
  \label{eq:ZNA}.
\eeq

By using the symmetry $G$, any 
vacuum can be transformed into the form 
\beq
 \left< \Phi \right> = v {\bf 1}_N,\quad
 v^2 = \frac{\mu^2}{2(N\lambda_1+\lambda_2)} . 
\eeq
We can consider this vacuum without loss of generality. 
All other degenerate vacua are obtained from this 
by the $G$ transformations.
The symmetry $G$ is spontaneously broken down to 
the isotropy group
\beq
 H = {SU(N)_{\rm V} \times ({\mathbb Z}_N)_{\rm A} 
      \over {\mathbb Z}_N \times {\mathbb Z}_N} 
   = {SU(N)_{\rm V}  
      \over ({\mathbb Z}_N)_{\rm V} }.
  \label{eq:isotropy}
\eeq
where ${\mathbb Z}_N \times {\mathbb Z}_N$ 
is the one of Eq.~(\ref{eq:ZNxZN}) which can be rewritten 
by Eqs.~(\ref{eq:ZNV}) and (\ref{eq:ZNA}),   
and $SU(N)_{\rm V}$ is given by
\beq
SU(N)_{\rm V}:&& (1,g,g^\dagger)
    \in U(1)_{\rm A} \times SU(N)_{\rm L} \times SU(N)_{\rm R} .
\eeq 
When the vacuum $\left< \Phi \right>$ is transformed by $G$, 
the isotropy group $H$ is also transformed by a similarity transformation, 
but all of them are isomorphic to 
the original isotropy group (\ref{eq:isotropy}).
Therefore the vacuum manifold 
(the order parameter space) 
can be written as a coset space:
\beq
{G \over H} 
 &=& {(U(1)_{\rm A} \times SU(N)_{\rm L} \times SU(N)_{\rm R})
    /({\mathbb Z}_N \times {\mathbb Z}_N)
  \over SU(N)_{\rm V}/({\mathbb Z}_N)_{\rm V} }  \non
 &\simeq&
  \frac{U(1)_{\rm A} \times SU(N)_{\rm A}}{({\mathbb Z}_N)_{\rm A}} 
 \simeq U(N)_{\rm A} , 
  \label{eq:U(N)A}
\eeq 
which is eventually a group manifold 
(because $H$ is a normal subgroup of $G$).
Since the first homotopy group of the vacuum manifold 
\beq
 \pi_1 [U(N)_{\rm A}] \simeq {{\mathbb Z}} 
\eeq
is non-trivial, it admits topological vortex-string solutions.

In order to find the mass spectrum, let us perturb $\Phi$ with small fluctuations as
\beq
\Phi(x) = v {\bf 1}_N + \phi (x) {\bf 1}_N + \chi_a (x) T^a,\quad
(a = 1,2,\cdots,N^2-1),
\eeq
with the generators $T^a$ of $SU(N)$ ($\Tr[T^aT^b]= \delta^{ab}$).
Then it is turned out that imaginary parts of $\phi$ and $\chi_a$ 
are massless Nambu-Goldstone modes 
parametrizing the vacuum manifold $U(N)_{\rm A}$ in Eq.~(\ref{eq:U(N)A}), 
while their real parts are massive:
\beq
m_\phi^2 = 2\mu^2,\quad
m_\chi^2 = \frac{2 \lambda_2}{N\lambda_1+\lambda_2} \mu^2 = 4 \lambda_2 v^2.
\eeq
The original coupling constants are expressed by these masses as
\beq
\mu^2 = \frac{m_\phi^2}{2},\quad
\lambda_1 = \frac{m_\phi^2 - m_\chi^2}{4Nv^2},\quad
\lambda_2 = \frac{m_\chi^2}{4v^2}.
\eeq 
The scalar potential can be rewritten by the dimensionfull
parameters $v,m_{\phi}$ and $m_\chi$ as
\beq
V = \frac{m_\phi^2}{4 N v^2} 
\left(\Tr\left[\Phi^\dagger\Phi - v^2 {\bf 1}_N\right]\right)^2 + 
\frac{m_\chi^2}{4v^2} 
\Tr\left[\left<\Phi^\dagger\Phi\right>^2\right],
\label{eq:pot2}
\eeq
where we have introduced a notation $\left<X\right> \equiv X - \frac{\Tr[X]}{N}{\bf 1}_N$ for $N\times N$ matrix $X$
(note $\left<{\bf 1}_N\right> = 0$).  
Let us introduce 
a ratio of the two masses\footnote{Instead of this, 
$\kappa = \lambda_1/\lambda_2$ was used to parametrize solutions  in \cite{Nitta:2007dp}.
}
\beq
\tau \equiv \frac{m_\chi}{m_\phi} = \sqrt{\frac{\lambda_2}{N\lambda_1 + \lambda_2}}.
\eeq
We will see that $\tau$ determines all properties of non-Abelian global vortices.

Let us construct vortex-string solutions in this model. 
Firstly, one may consider the following simple boundary condition
\beq
 \lim_{r \to \infty}\Phi(r,\theta) 
 = v e^{i\theta} {\bf 1}_N . 
\eeq
A natural ansatz with this boundary condition is 
\beq
 \Phi(r,\theta)  = ve^{i\theta} f(r) {\bf 1}_N,
 \label{eq:ans2}
\eeq
with boundary conditions $f(0) = 0$ and $f(\infty) = 1$.
The equation of motion for the profile function $f(r)$ is 
identical to that of the global $U(1)$ vortex with $k=1$ 
\beq
f'' + \frac{f'}{r}-\frac{f}{r^2}-\frac{m_\phi^2}{2} f \left(f^2 - 1\right) = 0.
\label{eq:abelianvortex}
\eeq
So the solution is just embedding of 
the global $U(1)$-vortex of Eq.~(\ref{eq:eom_gv}).
It is important to observe that neither $N$ nor $m_\chi$ appears. 
Furthermore, only the overall $U(1)$ phase winds once 
when we go around the vortex solution.
This Abelian vortex solution is called the $\eta'$ string 
\cite{Brandenberger:1998ew}.
However, as we  will see, this solution is not 
minimal \cite{Nitta:2007dp} and is broken 
into $N$ minimal solutions \cite{Nakano:2007dq}.

A minimal winding vortex configuration can be obtained 
by taking the following boundary condition
\beq
\lim_{r \to \infty}\Phi(r,\theta) =
 v\, {\rm diag}(e^{i\theta},1,\cdots,1) = 
v\,e^{i\frac{\theta}{N}} 
{\rm diag}\left(e^{i\frac{N-1}{N}\theta},e^{-i\frac{\theta}{N}},\cdots,e^{-i\frac{\theta}{N}}\right) .
\eeq
The key point here is that the overall $U(1)$-phase winds only 
$2\pi/N$ around the vortex and
the rests $2\pi (N-1)/N$ or $-2\pi/N$ are 
inside the non-Abelian group $SU(N)_{\rm A}$. 
As a consequence, the tension of the
vortex is $1/N$ of the above Abelian solution (\ref{eq:ans2}). Because the vacuum space is $U(N)_{\rm A}$,
the above vortex is called a non-Abelian global vortex 
or more specifically a global $U(N)$ vortex.
Corresponding vortex ansatz is of the form
\beq
\Phi(r,\theta)  = v\, {\rm diag}\left(e^{i\theta} f(r),g(r),\cdots,g(r)\right),
 \label{eq:ansat_nagv}
\eeq
with the boundary conditions
\beq
\lim_{r \to \infty} f = \lim_{r \to \infty} g = 1,\quad
\lim_{r \to 0} f = 0,\quad
\lim_{r \to 0} g' = 0. 
  \label{eq:boundary-cond}
\eeq
Since $f$ is the profile function of a winding scalar field, it must go to zero at the origin.
On the other hand, $g$ does not have to vanish.

For a fixed $\theta = \theta_0$, 
the field $\Phi$ behaves at the boundary 
as 
$\Phi(\theta=\theta_0, r \to \infty) = v\,{\rm diag}(e^{i\theta_0},1,\cdots,1)$ where the isotropy group is not the same with the one in Eq.~(\ref{eq:isotropy}). 
Instead, it is obtained from $H$ in Eq.~(\ref{eq:isotropy}) by a $G$ transformation 
$(e^{i\frac{\theta_0}{N}},g_0,g_0) \in U(1)_{\rm A} \times SU(N)_{\rm L }\times SU(N)_{\rm R}$ with 
$g_0 = {\rm diag}\left(e^{i\frac{N-1}{2N}\theta_0},e^{-i\frac{\theta_0}{2N}},\cdots,e^{-i\frac{\theta_0}{2N}}\right)$, 
and therefore it is isomorphic to $H=SU(N)_{\rm V}/({\mathbb Z}_N)_{\rm V}$ in Eq.~(\ref{eq:isotropy}).  
Around the vortex  where $g(r)$ differs from $f(r)$  
(especially at the center 
$\Phi(0) = v\,{\rm diag}(0,g(0),\cdots,g(0))$), 
the isotropy group 
$H_{\theta=\theta_0} = SU(N)_{{\rm V},\theta_0}/({\mathbb Z}_N)_{{\rm V},\theta_0}$ is further 
broken to its subgroup 
$[SU(N-1)_{\rm V} \times U(1)_{\rm V}]/({\mathbb Z}_N)_{{\rm V},\theta_0}$.  
This gives rises to further Nambu-Goldstone modes \cite{Nitta:2007dp}
\beq 
 {\mathbb C}P^{N-1} 
 \simeq {SU(N)_{\rm V} \over SU(N-1)_{\rm V} \times U(1)_{\rm V}}. 
  \label{eq:orientation}
\eeq 
In the case of the local $U(N)$ vortices for which 
$U(1)$ and $SU(N)_{\rm L}$ are gauged \cite{Hanany:2003hp}, 
the corresponding modes are called the orientational zero modes of a vortex. 
However, these are non-normalizable for a global $U(N)$ vortex 
(in the infinite space), 
unlike the local $U(N)$ vortex, 
because the isotropy groups $H_{\theta}$ depends on $\theta=\theta_0$ 
and differ from each other, and consequently  
the wave functions of the Nambu-Goldstone modes 
${\mathbb C}P^{N-1}$ spread to infinity.  

Let us investigate concrete profile functions 
of vortex solutions.
The Hamiltonian density and the energy densities 
in terms of the profile functions $f(r)$ and $g(r)$ is given by
\beq
{\cal H} &=& 2\pi v^2 r\, {\cal E},\\
{\cal E} &=& {f'}^2+ \frac{f^2}{r^2} + (N-1) {g'}^2 + {\cal V}, \\
{\cal V} &=& \frac{m_{\phi }^2}{4N} \left(f^2+(N-1) g^2-N\right)^2 
+ \frac{(N-1)m_{\chi }^2}{4 N} \left(f^2-g^2\right)^2 .
\eeq
By minimizing the Hamiltonian ${\cal H}$, 
we get the equations of motion for $f$ and $g$
\beq
&&f'' + \frac{f'}{r}-\frac{f}{r^2}-\frac{m_{\phi }^2}{2 N} f \left(f^2+(N-1) g^2 - N\right) 
-\frac{(N-1) m_{\chi }^2}{2 N} f \left(f^2-g^2\right) = 0,
\label{eq:nagv1}\\
&&g'' + \frac{g'}{r} - \frac{m_{\phi }^2}{2 N} g \left(f^2+(N-1) g^2 - N \right) 
+ \frac{m_{\chi }^2}{2N} g \left(f^2-g^2\right)  = 0.
\label{eq:nagv2}
\eeq
The third term in the left hand side of Eq.~(\ref{eq:nagv1}) is typical for global vortices.
This leads to logarithmic energy divergence.  
With respect to a dimensionless coordinate $\rho = m_\phi r$, the above equations can be 
written in the following forms
\beq
{\cal F}_N[f,g;\tau] \!\!&\equiv&\!\!
f'' + \frac{f'}{\rho} - \frac{f}{\rho^2} - \frac{f\left(f^2+(N-1) g^2 - N\right)}{2 N}  
-\frac{(N-1) \tau^2 f\left(f^2-g^2\right)}{2 N}   = 0,
\label{eq:nagv1a} \\
{\cal G}_N[f,g;\tau] \!\!&\equiv&\!\!
g'' + \frac{g'}{\rho} - \frac{g}{2 N} \left(f^2+(N-1) g^2 - N \right) 
+ \frac{\tau^2 g }{2N} \left(f^2-g^2\right)  = 0.
\label{eq:nagv2a}
\eeq
It is clear in this form that solutions depend on $\tau$ only. 
We should solve these ordinary differential
equations with boundary conditions (\ref{eq:boundary-cond}) 
with replacing $r$ by $\rho$.

Let us see the non-Abelian vortex in a special case $\tau=1$. 
In this case, 
${\cal G}_N[f,g;1]=0$ can be solved by $g = 1$ while 
the other equation ${\cal F}_N[f,g=1;\tau=1]=0$ 
is the same as (\ref{eq:eom_gv}). 
Therefore in the case of $\tau=1$ the Abelian vortex 
is embedded into $\Phi$ as a non-Abelian vortex solution.
Qualitative behaviors of vortex profiles change at $\tau=1$.
%%%%%%%%%%%%%%%%%%%%%%%%
\begin{figure}[ht]
\begin{center}
\begin{tabular}{cc}
\includegraphics[width=8cm]{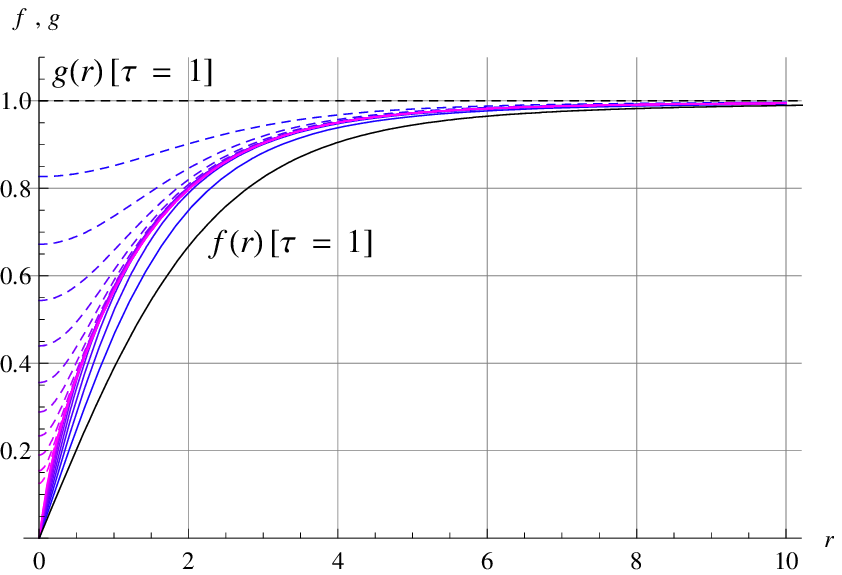} & \includegraphics[width=8cm]{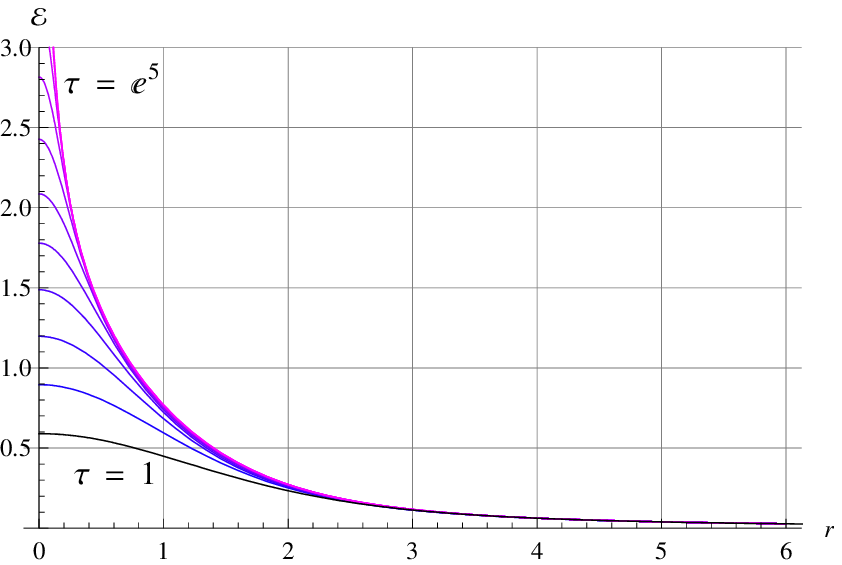}
\end{tabular}
\caption{Configurations (solid lines for $f(r)$ and broken lines for $g(r)$)
in the left panel and energy densities in the right panel for 
$\log \tau =0,1/2,1,3/2,\cdots,9/2,5$ for $N=2$.}
\label{fig:N=2_tg1}
\end{center}
\end{figure}
\begin{figure}[ht]
\begin{center}
\begin{tabular}{cc}
\includegraphics[width=8cm]{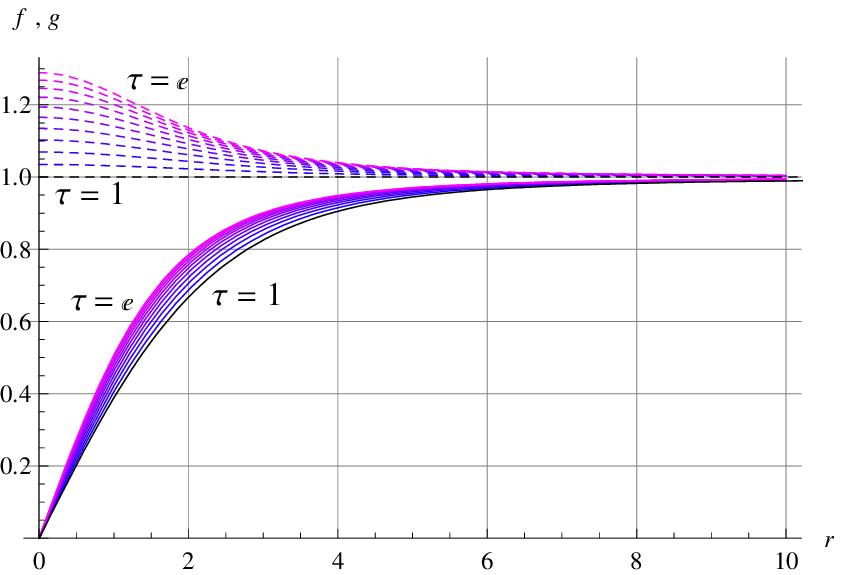} & \includegraphics[width=8cm]{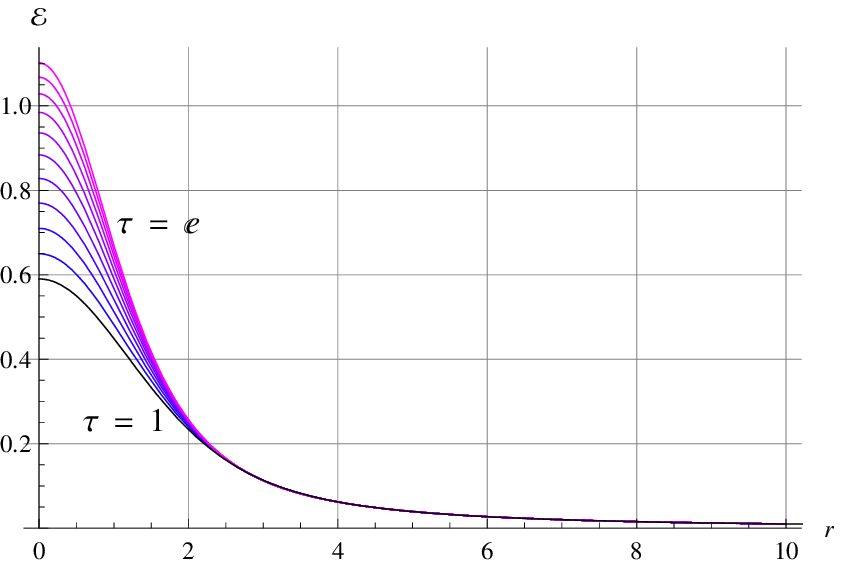}
\end{tabular}
\caption{Configurations (solid lines for $f(r)$ and broken lines for $g(r)$)
in the left panel and energy densities in the right panel for 
$- \log \tau =0,1/10,1/5,\cdots,9/10,1$ with $N=2$.}
\label{fig:N=2_tl1}
\end{center}
\end{figure}
%%%%%%%%%%%%%%%%%%%%%%%%%%
As we take $\tau$ to larger than 1, the value $g(0)$ gradually goes down toward zero and
the energy density becomes sharp (but remains regular for finite $\tau$), see Fig.~\ref{fig:N=2_tg1}. 
On the other hand, $g(0)$ grows and 
the vortex remains regular and finite when
we take $\tau$ smaller than 1, see Fig.~\ref{fig:N=2_tl1}. 
As we will see below, $g(0)$ is an important value.

\begin{figure}[ht]
\begin{center}
\begin{tabular}{cc}
\includegraphics[width=8cm]{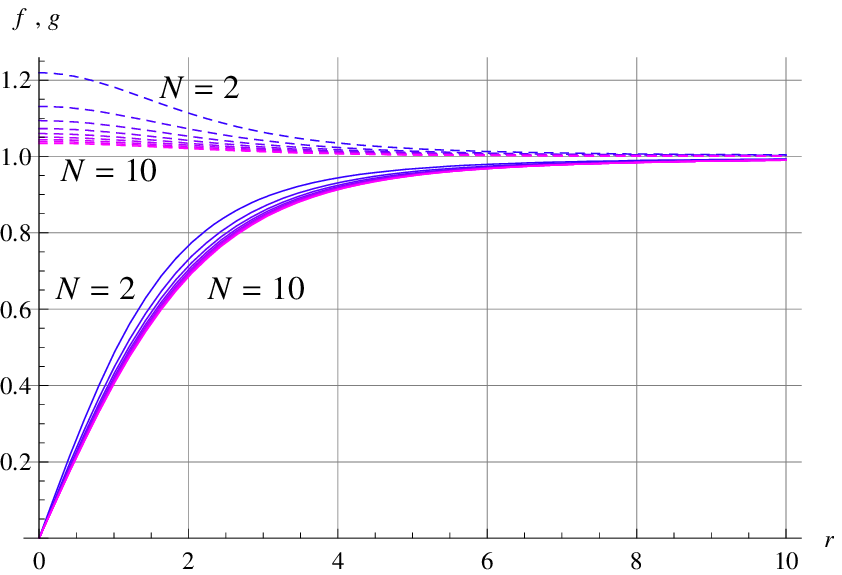} & \includegraphics[width=8cm]{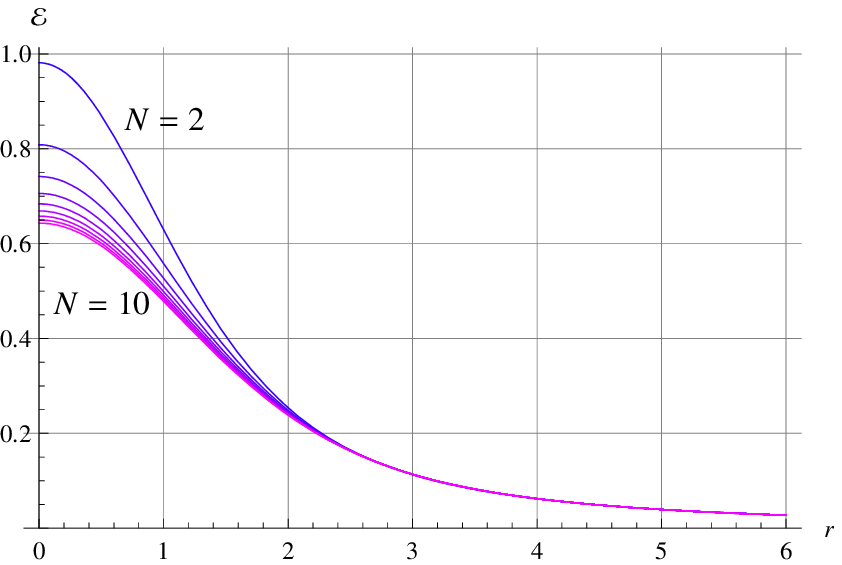}
\end{tabular}\\
\begin{tabular}{cc}
\includegraphics[width=8cm]{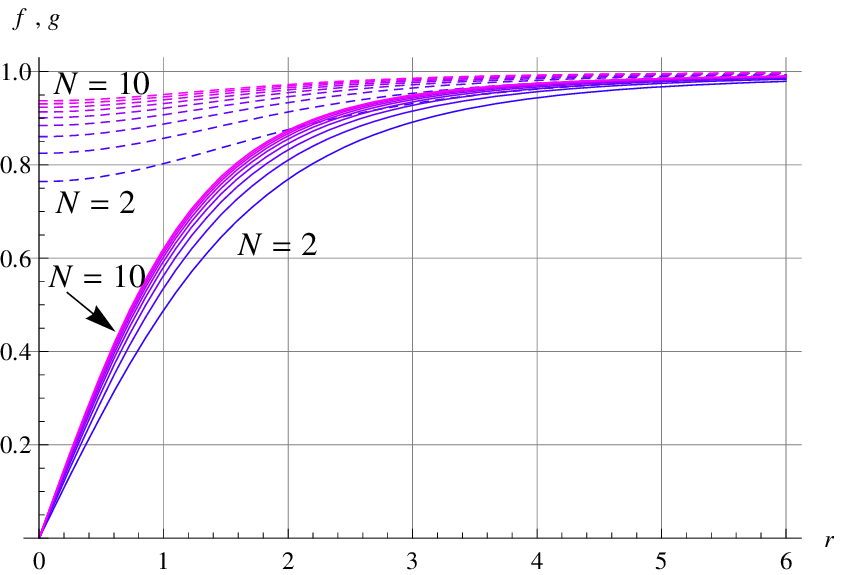} & \includegraphics[width=8cm]{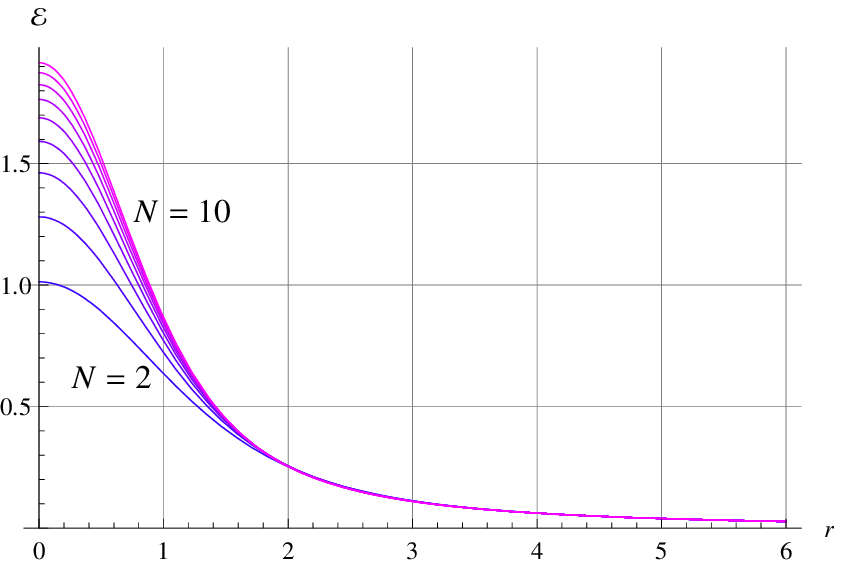}
\end{tabular}
\caption{
Configurations (solid lines for $f(r)$ and broken lines for $g(r)$) 
in the left two panels and energy densities 
in the right two panels, with $N=2,3,\cdots,10$. 
The upper(lower) two figures are for $\tau=1/2$($\tau = 2$).}
\label{fig:nh}
\end{center}
\end{figure}

In order to see $N$ dependence of the non-Abelian global vortex, we plot two figures in
Fig.~\ref{fig:nh} by changing $N$ form 2 to 10 with $\tau$ being fixed to $1/2$ and $2$.
%%%%%%%%%%%%%%%%%%%%%%%%%%
%%%%%%%%%%%%%%%%%%%%%%%%%%%%

%%%%%%%%%%%%%%%%%%%%%%%%%%%%%%%%%%%%%
\subsection{Asymptotic behaviors}

Let us investigate asymptotics of the non-Abelian global vortex.
This has been studied in \cite{Nitta:2007dp} and here we want to extend it
with a better accuracy and determine some fundamental parameters (shooting parameters)
accompanied by the differential equations (\ref{eq:nagv1a}) and (\ref{eq:nagv2a}).

We start with study on asymptotics at $r \gg \max\{m_\phi^{-1},m_\chi^{-1}\}$.
As usual, we expand the fields $f,g$ in the following
\beq
f = 1 - \sum_{i=1}^\infty a_i r^{-i},\quad
g = 1 - \sum_{i=1}^\infty b_i r^{-i}.
\eeq
By plugging these into Eqs.~(\ref{eq:nagv1a}) and (\ref{eq:nagv2a}) and equating them with 0 order by order,
we can analytically determine the coefficients. 
It is immediately turned out that
$a_{2i+1}$ and $b_{2i+1}$ vanish. 
The leading order terms are of the form
\beq
f = 1 - \frac{1}{N r^2} \left(\frac{1}{m_{\phi }^2} + \frac{N-1}{m_{\chi }^2} \right), \quad 
g = 1 - \frac{1}{N r^2}\left(\frac{1}{m_{\phi }^2}-\frac{1}{m_{\chi }^2}\right).
\eeq
Note that $f$ is always less than 1 
while $g \gtreqless 1$ for $1 \gtreqless \tau$.
As observed in \cite{Nitta:2007dp}, the behavior of $g(r)$ depends on the coupling constants $\lambda_1,\lambda_2$.
With respect to the physical masses $m_\chi$ and $m_\phi$, we now understand that the behavior depends on the ratio $\tau$
of the two masses.
The higher order terms are determined as
\beq
a_4 &=& \frac{(-1+N) (-1+9 N) m_{\phi }^4+2 (-1+N) m_{\phi }^2 m_{\chi }^2
+(1+8 N) m_{\chi }^4}{2 N^2 m_{\phi }^4 m_{\chi }^4},\\
b_4 &=& \frac{(1-8 N) m_{\phi }^4-2 m_{\phi }^2 m_{\chi }^2+(1+8 N) m_{\chi }^4}{2 N^2 m_{\phi }^4 m_{\chi }^4},\\
a_6 &=& \frac{1
}{2 N^3 m_{\phi }^6 m_{\chi }^6}
\big(
(N-1) (1+N (-58+161 N)) m_{\phi }^6 + (3-(86-83 N) N) m_{\phi }^4 m_{\chi }^2 \nonumber\\
&& - (3+(5-8 N) N) m_{\phi }^2 m_{\chi }^4 + (1+32 N (1+4 N)) m_{\chi }^6
\big),\\
b_6 &=& 
\frac{m_{\chi }^2-m_{\phi }^2}{2 N^3 m_{\phi }^6 m_{\chi }^6}
\big(\left(1-56 N+152 N^2\right) m_{\phi }^4+2 \left(-1+12 N+64 N^2\right) m_{\phi }^2 m_{\chi }^2 \nonumber\\
&& +\left(1+32 N+128 N^2\right) m_{\chi }^4\big),\\
&\vdots& \nonumber
\eeq

Let us next consider asymptotics at $r \ll \min\{m_\phi^{-1},m_\chi^{-1}\}$.
We expand the fields by
\beq
f = \sum_{i=0}^\infty c_ir^i,\quad
g = \sum_{i=0}^\infty d_ir^i.
\eeq
By plugging this into Eqs.~(\ref{eq:nagv1a}) and (\ref{eq:nagv2a}), one can determine
$c_{2m} = d_{2m+1} = 0$.
The leading order of the approximation gives us 
\beq
f = c_1 r + {\cal O}(r^3),\qquad g = d_0 + {\cal O}(r^2).
\eeq
The two parameters $c_1 = f'(0)$ and $d_0 = g(0)$ cannot be obtained analytically, but we can do it numerically.
Note that they depend on $\tau$ and $N$, see Table \ref{table:data_nagv}.
\begin{table}[ht]
\begin{center}
\begin{tabular}{c|cc|cc|cc|cc}
 & \multicolumn{2}{c|}{$N=2$} & \multicolumn{2}{c|}{$N=3$} 
& \multicolumn{2}{c|}{$N=4$} & \multicolumn{2}{c}{$N=5$} \\
\hline
$\log \tau$ & $c_1$ & $d_0$ & $c_1$ & $d_0$ & $c_1$ & $d_0$ & $c_1$ & $d_0$ \\
\hline
$+ \infty$ & -- & 0 & -- & 0 & -- & 0 & -- & 0  \\
3 & 0.89928 & 0.28858 & 1.35826 & 0.36708 & 1.77660 & 0.42829 & 2.15432 & 0.47785  \\
5/2 & 0.82145 & 0.35595 & 1.17489 & 0.44008 & 1.47803 & 0.5032 & 1.73919 & 0.55277  \\
2 & 0.74680 & 0.43959 & 1.00794 & 0.52694 & 1.21660 & 0.58945 & 1.38688 & 0.63682  \\
3/2 & 0.67273 & 0.54356 & 0.85220 & 0.62924 & 0.98383 & 0.68702 & 1.08462 & 0.72889  \\
1 & 0.59541 & 0.67207 & 0.70243 & 0.74686 & 0.77302 & 0.79353 & 0.82316 & 0.82555  \\
1/2 & 0.50986 & 0.82689 & 0.55511 & 0.87497 & 0.58130 & 0.90209 & 0.59840 & 0.91953 \\
0 & 0.41238 & 1 & 0.41238 & 1 & 0.41238 & 1 & 0.41238 & 1  \\
$-1/2$ & 0.50651 & 1.16517 & 0.46970 & 1.10162 & 0.45361 & 1.07342 & 0.44457 & 1.05748  \\
$-1$ & 0.57045 & 1.28860 & 0.50278 & 1.16697 & 0.47578 & 1.11765 & 0.46121 & 1.09086  \\
$-3/2$ & 0.60330 & 1.35937 & 0.51809 & 1.20061 & 0.48563 & 1.13948 & 0.46845 & 1.10697  \\
$-2$ & 0.61733 & 1.39235 & 0.52430 & 1.21534 & 0.48955 & 1.14881 & 0.47131 & 1.11377  \\
$-5/2$ & 0.62280 & 1.40589 & 0.52667 & 1.22120 & 0.49104 & 1.15249 & 0.47239 & 1.11644  \\
$-3$ & 0.62486 & 1.41111 & 0.52755 & 1.22343 & 0.49160 & 1.15388 & 0.47279 & 1.11744 \\
$- \infty$ & 0.62607 & $\sqrt 2$ & 0.52807 & $\sqrt{3/2}$ &  0.49192 & $\sqrt{4/3}$  & 0.47302 & $\sqrt{5/4}$ \\
\hline
\end{tabular}
\caption{Numerical data for the shooting parameters. 
The parameter $d_0$ at $\log \tau \to - \infty (\tau \to 0)$ has 
analytic values, given in Eq.~(\ref{eq:lim1}), below.
}
\label{table:data_nagv}
\end{center}
\end{table}

Higher order terms are determined by $c_1$ and $d_0$:
\beq
d_2 &=& - \frac{N \left(1-d_0^2\right) m_{\phi }^2 + \left(m_{\phi }^2-m_{\chi }^2\right)  d_0^2}{8N}\,d_0 ,\\
c_3 &=& \frac{-N m_{\phi }^2 +(N-1)  \left(m_{\phi }^2-m_{\chi }^2\right)d_0^2}{16 N}\,c_1,\\
d_4 &=& \frac{d_0}{256 N^2} \bigg[8 N c_1^2 \left(m_{\phi }^2-m_{\chi }^2\right)\nonumber\\
&& +\left(N \left(3 d_0^2-1\right) m_{\phi }^2-3 d_0^2 \left(m_{\phi }^2-m_{\chi }^2\right)\right) 
\left(N \left(d_0^2-1\right) m_{\phi }^2-d_0^2 \left(m_{\phi }^2-m_{\chi }^2\right)\right)\bigg],\\
c_5 &=& \frac{c_1}{768 N^2} \bigg[\left(N^2-6 (N-1) N d_0^2+5 (N-1)^2 d_0^4\right) m_{\phi }^4 \nonumber\\
&&-2 (N-1) d_0^2 \left(-3 N+(-5+3 N) d_0^2\right) m_{\phi }^2 m_{\chi }^2
+(5+(N-6+) N) d_0^4 m_{\chi }^4 \nonumber\\
&& +16 N c_1^2 \left(m_{\phi }^2+(N-1) m_{\chi }^2\right)\bigg]. 
\eeq

%%%%%%%%%%%%%%%%%%%%%%%%%%%%%%%%%%%%
\subsection{Heavy Particle Limits}

In this subsection, we are going to study how the vortices behave in two regions
i) $m_\phi \gg m_\chi$ and ii) $m_\chi \gg m_\phi$. Before doing it, remember that the
linear sigma model (\ref{eq:lsm}) becomes the chiral Lagrangian in the limit
$m_\chi,m_\phi \to \infty$. This is because, after the heavy fields are integrated out,
only massless NG modes ($\Phi \in U(N)$) survive in a low energy theory. 
Our purpose of this subsection is to clarify effects of lightest massive fields to the chiral Lagrangian.
For later convenience, let us rewrite the equations of motion (\ref{eq:nagv1})
and (\ref{eq:nagv2}) as follows
\beq
&&\frac{f''}{f} + \frac{f'}{rf}-\frac{1}{r^2}
+ (N-1)\left(\frac{g''}{g} + \frac{g'}{rg}\right)
-\frac{m_{\phi }^2}{2 } \left(f^2+(N-1) g^2 - N\right) = 0,
\label{eq:nagv1b}\\
&& \frac{f''}{f} + \frac{f'}{rf}-\frac{1}{r^2} - \frac{g''}{g} - \frac{g'}{rg}
- \frac{m_{\chi }^2}{2} \left(f^2-g^2\right)  = 0.
\label{eq:nagv2b}
\eeq
A point is that $m_\phi$ appears only in the first equation while 
$m_\chi$ is in the second equation.

Similar limits (with some particles being infinitely heavy) 
have been studied recently 
for a {\it local} $U(N)$ vortex,  
for which $U(1)$ and $SU(N)_{\rm L}$ are gauged, 
at the critical coupling 
(with a certain relation between gauge and scalar couplings) \cite{efnnos}.

%%%%%%%%%%%%%%%%%%%%%%%%%%%%%%%%%%%%%%%%%%%%%%
\subsubsection{$m_\phi \to \infty$ limit $(\tau \to 0)$}

Let us first consider the mass of the trace part of $\Phi$ is much greater than that of traceless part,
\beq
m_\phi \gg m_\chi.
\eeq
In other words, we consider a limit $\tau \to 0$ by sending $m_\phi \to \infty$ with
$m_\chi$ being fixed\footnote{
The same limit can be realized by taking $m_\chi \to 0$ with $m_\phi$ being fixed.
However, the massless limit is tricky because the vortex infinitely spreads out and dilutes.
}.
Then we integrated out the heavy modes with mass $m_\phi$.
The first term in the scalar potential in Eq.~(\ref{eq:pot2}) becomes very high and sharp, so
that the scalar fields are squeezed into the following manifold
\beq
\Tr[\Phi^\dagger\Phi - v^2 {\bf 1}_N] = 0.
\eeq
To solve this condition, let us expand $\Phi$ by
\beq
\Phi = \vec \varphi \cdot \vec T,\quad
\vec T = ({\bf 1}_N/{\sqrt N},T^1,T^2,\cdots,T^{N^2-1}),
\eeq
with $N^2$ complex vector $\vec \varphi = (\varphi_0,\varphi_1,\cdots,\varphi_{N^2-1})$. 
Then the condition becomes
\beq
|\vec \varphi|^2 = Nv^2. \label{eq:cond}
\eeq
Therefore the linear sigma model (\ref{eq:lsm}) reduces to $S^{2N^2-1}$ non-linear sigma model.
The scalar potential takes the form
\beq
V_0 &=& \frac{m_\chi^2}{4v^2}\Tr\left[
\left<\Phi^\dagger\Phi\right>^2
\right] 
= \frac{m_\chi^2}{4v^2}\Tr\left[
\left(\Phi^\dagger\Phi - v^2 {\bf 1}_N\right)^2
\right].
\eeq
The vacuum remains as $U(N)_{\rm A}$. 
By using the chiral symmetry, we can choose
\beq
\varphi_0 = \sqrt N v,\quad \varphi_{a\ge1} = 0
\quad\Leftrightarrow\quad
\Phi = v {\bf 1}_N.
\eeq

Let us next consider a vortex-string solution in this model by taking the same 
diagonal ansatz (\ref{eq:ansat_nagv}) as before. 
However, $f$ and $g$ are no longer independent because of 
the condition (\ref{eq:cond}).
They should satisfy
\beq
f(r)^2 + (N-1)g(r)^2 = N.
\label{eq:mp_inf}
\eeq
With this at hand, we are aware of an interesting phenomenon that
the shooting parameter $g(0) = d_0$ approaches to an analytic value as
\beq
\lim_{\tau \to 0} d_0 = \lim_{\tau \to 0} g(0) = \sqrt{\frac{N}{N-1}}.
\label{eq:lim1}
\eeq
Here we have used $f(0)=0$. Our numerical result matches with this result, see Table \ref{table:data_nagv}.
The same thing can be found from the view point of equations of motion
(\ref{eq:nagv1b}). By dividing its both hands by $m_\phi^2$ and taking the limit $m_\phi \to \infty$,
we get Eq.~(\ref{eq:mp_inf}) again. 

\begin{figure}[ht]
\begin{center}
\includegraphics[width=16cm]{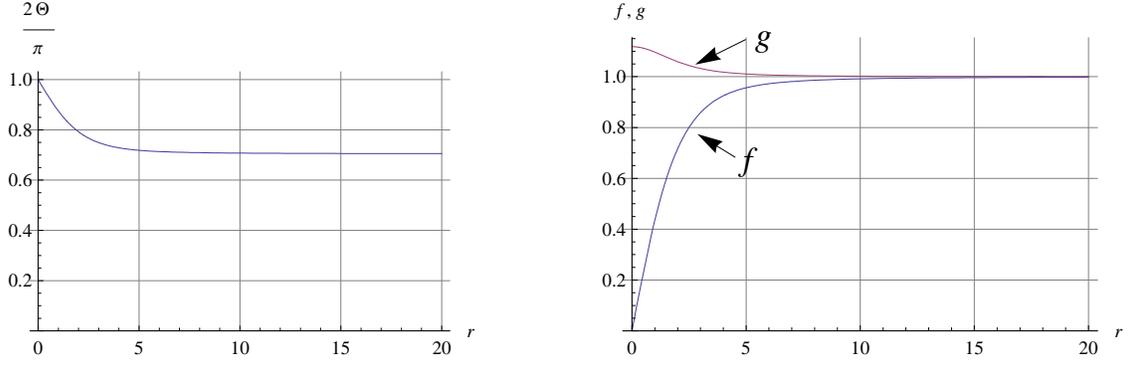} 
\caption{Profiles of $\Theta$ and corresponding $f,g$ for $N=2$ case ($m_\chi = 1$).}
\label{fig:tau_0}
\end{center}
\end{figure}

To simplify the other equation (\ref{eq:nagv2b}),
we may rewrite the fields by
\beq
f = \sqrt N \cos \Theta,\quad
g = \sqrt{\frac{N}{N-1}} \sin \Theta.
\eeq
The reduced model is like the Sine-Gordon model: the potential is
\beq
\tau=0:\quad V = 
m_{\chi }^2\frac{\left(N \cos^2\Theta -1 \right)^2 }{4 (N-1)}.
\eeq
The corresponding equation of motion is of the form
\beq
\tau=0:\quad \Theta ''+\frac{\Theta '}{r}+\frac{1}{4} \sin2 \Theta 
\left(\frac{2}{r^2}+ m_{\chi }^2\,\frac{N \left(N \cos^2\Theta - 1 \right)}{N-1}\right)
= 0.
\eeq
We solve this with the boundary conditions
\beq
\lim_{r \to 0} \Theta = \frac{\pi}{2},\quad
\lim_{r \to \infty} \Theta = \arccos\left({\frac{1}{\sqrt N}}\right).
\eeq
We numerically solved this and the solution is shown in Fig.~\ref{fig:tau_0}.

%%%%%%%%%%%%%%%%%%%%%%%%%%%%%%%%%%%%%%%%%%%
\subsubsection{$m_\chi \to \infty$ limit $(\tau \to \infty)$} \label{sec:lim2}

Next let us investigate the other limit 
\beq
m_\chi \gg m_\phi,\quad (\tau \to \infty).
\eeq
We send $m_\chi$ infinity with $m_\phi$ being kept finite.
Then $N^2-1$ real scalar fields in $\Phi$ become infinitely heavy and
we integrated them out from the theory.
The scalar potential (\ref{eq:pot2}) gives us the following condition
\beq
\Tr\left[\left<\Phi^\dagger\Phi\right>^2\right] = 0
\quad \Leftrightarrow \quad
\left<\Phi^\dagger\Phi\right> = 0
\quad \Leftrightarrow \quad
\Phi^\dagger\Phi \propto {\bf 1}_N.
\label{eq:mc_inf}
\eeq
We can solve this by
\beq
\Phi(x) = s(x)\, U(x) 
  \quad {\rm with} \quad s \in {\mathbb C},\ U\in SU(N)_{\rm A} .
\eeq
This decomposition is up to 
the ${\mathbb Z}_k$ identification 
$(s,U) \sim (\omega^k s,\omega^{-k} U)$ with 
$\omega = e^{2\pi i/N}\, (k=0,1,2,\cdots,N-1)$.
Thus the theory in this limit is 
the $SU(N)$ chiral Lagrangian coupled with a complex scalar field $s$.
The kinetic term is given by
\beq
 K_{\tau \to \infty} 
 &=& N |\p_\mu s|^2 + |s|^2\Tr(\p_\mu U \p^\mu U^\dagger) \non
 &=& N (\p_\mu \sigma)^2 
 + \sigma^2 \Tr(\p_\mu \hat U \p^\mu \hat U^\dagger)
\eeq
with $s \equiv \sigma e^{i \alpha}$ and 
$\hat U \equiv e^{i \alpha} U \in U(N) \simeq [SU(N) \times U(1)] /{\mathbb Z}_N$.
From the second expression, we see that 
the metric of the target space is a cone over $U(N)$.  
The scalar potential becomes in this limit as 
\beq
V_{\tau \to \infty} = 
\frac{m_\phi^2}{4v^2}\left(\Tr[\Phi^\dagger\Phi - v^2 {\bf 1}_N]\right)^2
= \frac{N^2 m_\phi^2}{4v^2}\left(|s|^2 - v^2\right)^2.
 \label{eq:potential}
\eeq
Thus the vacuum manifold remains to be $U(N)_{\rm A}$.
The vacuum expectation value $v$ of $|s|$ gives the pion decay constant 
$F_\pi^2 = 16 |s|^2$ of the $U(N)$ chiral Lagrangian.

\begin{figure}[ht]
\begin{center}
\begin{tabular}{cc}
\includegraphics[width=8cm]{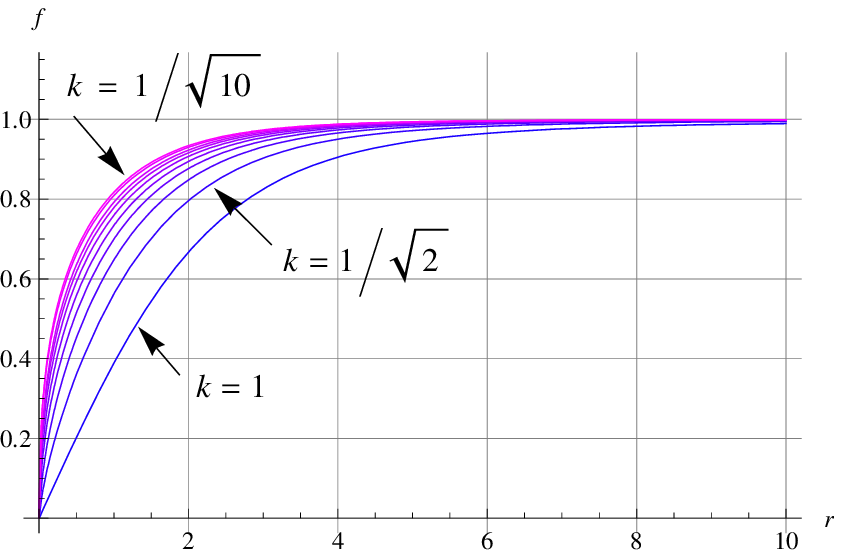} & 
\includegraphics[width=8cm]{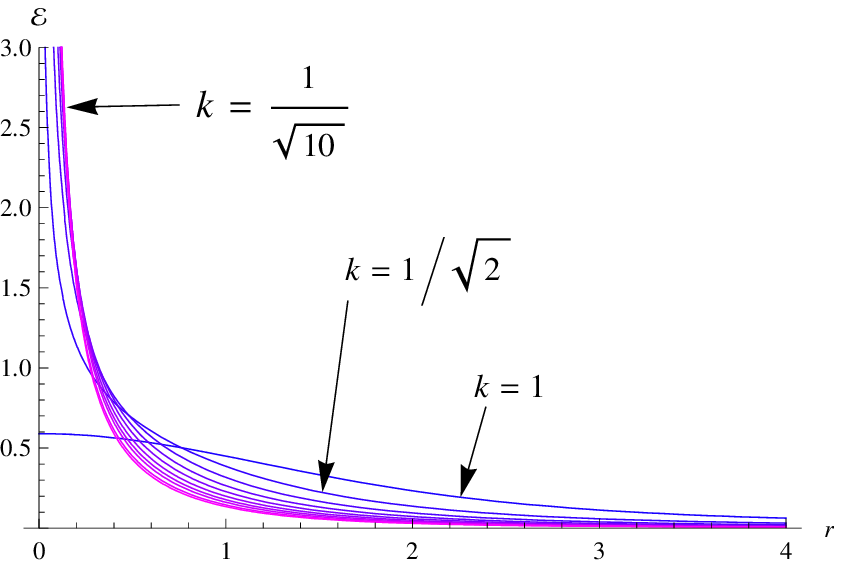} 
\end{tabular}
\caption{The left panel shows the profile functions 
and the corresponding energy densities ${\cal E}$ in Eq.~(\ref{eq:energy_u1}) are plotted in the right panel 
for $k=1/\sqrt{N}$ with $N=1,2,\cdots, 10$. 
The Abelian solution of $N=1$ is only regular but the others are singular.}
\label{fig:irr_gv}
\end{center}
\end{figure}

We are ready to consider vortex-string solutions in the limit.
With respect to the profile functions $f,g$, 
the condition (\ref{eq:mc_inf}) forces us the following condition
\beq
f(r) = g(r) . \label{eq:f=g} 
\eeq
In terms of $s$ and $U$, the non-Abelian vortex solution 
(\ref{eq:ansat_nagv}) can be expressed as 
\beq
s = e^{i\frac{\theta}{N}} f(r),\quad
U = {\rm diag}\left(e^{\frac{i(N-1)\theta}{N}},
e^{-i\frac{\theta}{N}},\cdots,e^{-i\frac{\theta}{N}}
\right). \label{eq:sU}
\eeq
Note that the overall $U(1)$ winding of the non-Abelian vortex is $1/N$ as before. 
The condition (\ref{eq:f=g}) explains the behavior of $g(0)$ 
which tends to go down toward zero as $\tau$ is sent to $\infty$,
see Fig.~\ref{fig:N=2_tg1} and Table \ref{table:data_nagv}.
Because of this behavior, the scalar fields $\Phi$ vanish 
at the origin entirely so that 
the full chiral symmetry is recovered at the center of vortex.
Furthermore with recalling the discussion around Eq.~(\ref{eq:orientation}), 
we find that no second symmetry breaking occurs 
in the presence of the vortex solution (\ref{eq:sU}), 
because the isotropy group is $H_{\rm \theta}$ even for finite $r (\neq 0)$   
which is isomorphic to 
$H=SU(N)_{\rm V}/({\mathbb Z}_N)_{\rm V}$ in Eq.~(\ref{eq:isotropy}).  
Therefore, we find that the solution loses 
the internal orientations of ${\mathbb C}P^{N-1}$. 
Eq.~(\ref{eq:nagv2b}) is automatically satisfied while
Eq.~(\ref{eq:nagv1b}) reduces to
\beq
\tau = \infty:\quad
f'' + \frac{f'}{r} - \frac{1}{Nr^2} - \frac{m_\phi^2}{2}f(f^2-1) = 0.
\label{chilargelimit} 
\eeq
Let us compare this with Eq.~(\ref{eq:agv}) for the Abelian vortex string.
Interestingly, the non-Abelian global vortex 
in the $m_\chi \to \infty$ limit can be identified with
the Abelian global vortex with a non-integer $U(1)$ winding number 
\beq
k=\frac{1}{\sqrt N}, 
\eeq
which is smaller than unity, and can be an {\it irrational} numbers for 
$N$'s which are not able to be expressed by squared integers.
Of course, $U(1)$ vortices 
with a non-integer $U(1)$ winding number 
are generically singular 
because $\phi \sim v\, e^{i\theta/\sqrt{N}}$ 
(at $r \sim 0$) 
is not single valued.
Such singular solutions for $F[f(\rho),\frac{1}{\sqrt N}] = 0$ 
in Eq.~(\ref{eq:eom_gv})
with the boundary conditions (\ref{eq:bc}) are shown in Fig.~\ref{fig:irr_gv}.
%%%%%%%%%%%%%%%%%%%
%%%%%%%%%%%%%%%%%%%%

There is no smooth interpolation 
in the solutions between 
$m_{\chi} < \infty$ and 
$m_{\chi}=\infty$ in the following sense: 
Eqs.~(\ref{eq:nagv1b}) and (\ref{eq:nagv2b}) with finite $m_{\chi}$ 
always provide $f \sim r$ to leading order in $r \sim 0$ 
(see Table \ref{table:data_nagv}), 
and keep the energy density finite, 
while in the limit where $m_{\chi}=\infty$ 
the resultant Eq.~(\ref{chilargelimit}) gives $f \sim r^{\frac{1}{\sqrt{N}}}$ for $r \sim 0$ 
and energy density gets divergent as $r^{2\left(\frac{1}{\sqrt{N}}-1\right)}$ 
at the center of vortex. 
Furthermore, the profile function $g$ (in this limit $g=f$)
does not satisfy the original boundary condition $g'(0)=0$.
Instead, it is replaced by $g(0) = 0$. 

Although this singular solution is an artifact appearing only in the limit 
where we have discarded the heavy modes completely, 
it reasonably accounts for 
the fact that the energy profile of the non-Abelian vortex becomes very sharp and
finally looks singular when $m_\chi \gg m_\phi$, see Fig.~\ref{fig:N=2_tg1}.
Taking into account such heavy modes, the singularity is smeared.

%%%%%%%%%%%%%%%%%%%%%%%%%%%%%%%%%%%%%%%%%%%%%
\section{Conclusion and Discussion}\label{sec:conc}

In this article, we have investigated  
non-Abelian global vortices in 
$SU(N)_{\rm L}\times SU(N)_{\rm R}\times U(1)_{\rm A}$ linear sigma model
in detail. We push forward the analysis in \cite{Nitta:2007dp} and determined important numerical parameters
$c_1$ and $d_0$ which determines all the properties of the solutions. Furthermore, we have obtained
expansion formulae for asymptotics at large distance.
We have found that interesting two limits i) $m_\phi \gg m_\chi$, ii) $m_\chi \gg m_\phi$.
The original linear sigma model reduces $S^{2N^2-1}$ non-linear sigma model in the i) limit and
we have found a sort of non-Abelian global string solution there. 
In the second limit ii), 
we have obtained the $SU(N)$ chiral Lagrangian
coupled with a complex scalar field. 
We have also found a sort of Abelian global vortex solution.
It is a singular solution and can be identified with Abelian global 
vortex with an irrational $U(1)$ winding number $k=1/\sqrt N$.
\footnote{
We have also investigated the Large $N$ limit, 
and found that there exists a regular vortex string solution 
which has the same form with the usual $U(1)$ vortex Eq.~(\ref{eq:abelianvortex}) 
with winding number $k=1$ but with replacement $m_\phi \rightarrow m_\chi$.
See appendix for details.}

Here we give several discussions. 
The coupling between a global $U(1)$ string
and the $U(1)$ Nambu-Goldstone boson can be 
constructed by using a duality between a boson $\phi$
and a two-form field $B_{\mu\nu}$ \cite{Kalb:1974yc}. 
In the same way the coupling of a global $U(N)$ string
and the $U(N)$ Nambu-Goldstone bosons will be 
possible by using non-Abelian two-form \cite{NA-2form}.

In the presence of the $U(1)_{\rm A}$ axial anomaly 
a term 
$V_1 (\Phi,\Phi^\dagger) = c ( \det \Phi + \det \Phi^\dagger)$
is induced, 
which gives a sine-Gordon potential to the phase. 
Then a $U(N)$ vortex becomes a boundary of a domain wall 
\cite{Balachandran:2001qn,Balachandran:2002je}. 
In the presence of quark masses, 
a term $V_2 (\Phi,\Phi^\dagger) = \Tr [H (\Phi + \Phi^\dagger )]$ 
exists with $H$ a mass matrix.
It remains as a future problem to study 
detailed structure of vortex solutions 
in the presence of these terms 
because the authors in 
\cite{Balachandran:2001qn,Balachandran:2002je} 
assumed constant profiles.

In Subsec.~\ref{sec:lim2} we have simply sent the mass $m_{\chi}$ 
to infinity to obtain the $SU(N)$ chiral Lagrangian coupled 
with a complex scalar field $s$. However quantum mechanically we should 
integrate out the massive fields. 
This procedure generally induces higher derivative terms for 
remaining massless fields. 
The Skyrme term is such a term of the fourth order \cite{Skyrme:1961vq}. 
In our case with massless field $s$ we will obtain 
the effective Lagrangian of the form 
\beq 
 {\cal L}_{\rm eff.} 
&=& N |\p_\mu s|^2  
 + |s|^2 \Tr(\p_\mu U \p^\mu U^\dagger) 
 + {|s|^4\over e^2} \Tr([U^\dagger \p_\mu U,U^\dagger \p_\nu U]^2)
 - V(|s|^2) \non
&=& N (\p_\mu \sigma)^2  
 + \sigma^2 \Tr(\p_\mu \hat U \p^\mu \hat U^\dagger) 
 + {\sigma^4\over e^2} 
  \Tr([\hat U^\dagger \p_\mu \hat U,\hat U^\dagger \p_\nu \hat U]^2)
 - V(\sigma^2)  \label{eq:skyrme} 
\eeq
where $e$ is a parameter determined by an explicit calculation 
and $V$ is the potential in Eq.~(\ref{eq:potential}). 
Note that there is no fourth order term for $s$.
Because of the relation 
$\Phi^\dagger \p_{\mu}\Phi 
= s^* \p_\mu s {\bf 1}_N + |s|^2 U^\dagger \p_\mu U$,  
the possible fourth order term 
$[\Phi^\dagger \p_\mu \Phi, \Phi^\dagger \p_\nu \Phi]^2$  
reduces to the Skyrme-like term in the Lagrangian (\ref{eq:skyrme}).
When $s$ is fixed to the vacuum expectation value, 
the Lagrangian (\ref{eq:skyrme}) reduces to the $U(N)$ Skyrme model 
and admits 
the usual Skyrmion solution for $U(x) = \exp i [F(r) \sigma \cdot {\bf r}/|{\bf r}|]$.\footnote{
The Skyrme model admits a topologically unstable 
string solution \cite{Jackson:1988xk} 
which may be related to the pion string \cite{Zhang:1997is}.
} 
It is interesting to note that 
the abelian vortex $s \sim v e^{i \theta}$ does not interact 
with the Skyrmion while
a non-Abelian vortex does. 
It remains as a future problem to study 
interaction, scattering or absorption of 
baryons(Skyrmions) by non-Abelian strings.

In this paper we have studied $U(N)$ vortices. 
Local and semi-local vortices with different groups 
$[U(1) \times G]/{\mathbb Z}_{n_0}$, 
where $G$ is arbitrary group with the center ${n_0}$
\cite{Eto:2008yi}, have been studied recently. 
In this framework the $U(N)$ vortex corresponds to the case 
of $G=SU(N)$ with $n_0=N$. 
Global version of these vortices are also possible, 
especially the case of $G=SO$ is related to 
vortices in the B-phase of $^3$He superfluids.

%%%%%%%%%%%%%%%%%%%%%%%%%%%%%%%%%%%%%
\bigskip
Before closing this paper let us compare our global $U(N)$ vortices with 
other types of $U(N)$ vortices in the related models: 
1) semi-superfluid $U(N)$ vortices 
in high density QCD and 2) local $U(N)$ vortices. 
In these models, the group structure is completely the same 
with the global case in this paper. 
However the energetics/interactions 
of vortices and the (non-)normalizability of the zero modes 
are significantly different. 

1) 
In high density QCD it is expected that color superconductivity is 
realized. 
There, the color symmetry $SU(N)_{\rm C}$ and the flavor symmetry 
$SU(N)_{\rm F}$ (with $N=3$) as well as the baryon $U(1)_{\rm B}$ symmetry 
are spontaneously broken down 
to the color-flavor locked symmetry $SU(N)_{\rm C+F}$ 
apart from the discrete symmetries. 
The corresponding vacuum manifold 
$[SU(N) \times U(1)]/{\mathbb Z}_N \simeq U(N)$ 
is the same with that of the global $U(N)$ vortices. 
In this case the $SU(N)$ subgroup of the vacuum manifold 
$U(N)$ is gauged and therefore only one massless Nambu-Goldstone 
boson for the $U(1)_{\rm B}$ exists.
The $U(N)$ vortices here are called semi-superfluid vortices 
\cite{Balachandran:2005ev}. 
In the asymptotic behaviour of the scalar field of a 
$U(N)$ vortex, 
$\Phi \sim v\, {\rm diag}(e^{i\theta},1,\cdots,1) = 
v\,e^{i\frac{\theta}{N}} 
{\rm diag}\left(e^{i\frac{N-1}{N}\theta},e^{-i\frac{\theta}{N}},\cdots,e^{-i\frac{\theta}{N}}\right)$, 
the latter non-Abelian part can be eliminated by a gauge transformation 
$U(r,\theta) = 
{\rm diag}\left(e^{-i\frac{N-1}{N}  \theta F(r)},e^{i\frac{\theta}{N} F(r)},\cdots,e^{i\frac{\theta}{N} F(r)}\right)$ with an arbitrary 
function $F(r)$ satisfying the boundary conditions 
$F(r=0) =0$ and $F(r\to\infty) =1$.\footnote{
This transformation is well-defined because of the triviality 
of the first homotopy group: $\pi_1[SU(N)] =0$.
This transformation brings $\Phi$ to 
$\Phi \sim v\,e^{i\frac{\theta}{N}} {\bf 1}_N$. 
This property makes the orientational zero modes of ${\mathbb C}P^{N-1}$ 
to be normalizable and to become 
the moduli (collective coordinates) of the vortex \cite{Nakano:2007dr}. 
In other words, the isotropy groups $H_{\rm \theta}$ at the infinities
in the presence of a global $U(N)$ vortex were physically 
different and depend on $\theta$ (although they are isomorphic), 
whereas, in the $SU(N)$ gauged case, 
the isotropy group at the infinities in the presence of 
a semi-superfluid $U(N)$ vortex
is physically equivalent for any $\theta$. 
The $U(1)_{\rm B}$ is global and so the energy of a vortex 
remains logarithmically divergent. 
The asymptotic interaction between two semi-superfluid $U(N)$ vortices 
is also essentially the same with the one between $U(1)$ global vortices 
because of the above property \cite{Nakano:2007dr,Nakano:2008dc}. 
Therefore it gives the universal repulsion 
between separated vortices.
}

2) Next let us to compare the global $U(N)$ vortices studied in 
this paper with the local $U(N)$ vortices \cite{Hanany:2003hp}.
In this case too the symmetry breaking pattern is the same 
but a crucial difference is that the $U(1)$ symmetry is also gauged 
in addition to the color $SU(N)$, 
and therefore there remain no Nambu-Goldstone bosons. 
Namely the vacuum is the unique at the infinity even 
in the presence of a $U(N)$ vortex because 
the $U(1)$ symmetry is gauged . 
Nevertheless we cannot gauge transform the 
asymptotic behavior of the scalar fields from 
$\Phi \sim v\,e^{i\frac{\theta}{N}}{\bf 1}_N$ 
to ${\bf 1}_N$ because we can define 
no regular gauge transformation well-defined in the entire space 
because of non-triviality of the first homotopy group: $\pi_1[U(1)] \neq 0$.
Unlike the global $U(N)$ vortices or semi-superfluid $U(N)$ vortices,  
these local $U(N)$ vortices have finite energy because of the gauged $U(1)$. 
The ${\mathbb C}P^{N-1}$ zero modes are normalizable.  
At the critical (BPS) coupling with a particular relation 
between gauge and scalar couplings, 
there is no static force among multiple vortices, 
with allowing the multi-vortex moduli space 
\cite{Eto:2005yh,Eto:2006pg} 
(see \cite{Eto:2006mz} for the moduli spaces of 
local $U(N)$ vortices on a cylinder and a torus). 
Static interactions exist between vortices  
at non-critical (non-BPS) couplings. 
The force between two $U(N)$ vortices was shown 
to depend on both ${\mathbb C}P^{N-1}$ orientations 
and positions \cite{Auzzi:2007wj}. 

If we gauge all the symmetry, the vortices are those 
in quiver gauge theories \cite{Popov:2005ik}. 
In this case, the diagonal gauge symmetry remains unbroken, 
and the vortices do not have orientations 
in the internal space.  
The final possibility which was not studied so far is the case 
that only $U(1)$ is gauged.

%%%%%%%%% Acknowledgements %%%%%%%%%%%%%
\section*{Acknowledgements}
%%%%%%%%%%%%%%%%%%%%%%%%%%%%%%%%%%%%%%%%

This work is supported in part by Grant-in-Aid for 
Scientific Research from the Ministry of Education, 
Culture, Sports, Science and Technology, Japan No.20740141 (M.N.).
The work is also supported by the Research Fellowships of the Japan Society 
for the Promotion of Science for Research Abroad (M.E). 
E.N. gratefully acknowledges financial support from 
the Frankfurt Institute for Advanced Studies.

\appendix
%%%%%%%%%%%%%%%%%%%%%%%%%%%%%%%%%%%%%%%%%%%%%%%
\section{Large $N$ Limit}
In this appendix we will derive 
an asymptotic form of the vortex solution in the large $N$ limit. 
Before implementing the large $N$ limit, 
we have to know how the parameters in the Lagrangian scale in $N$.
From the observation of loop corrections 
(perturbation series in $\lambda_1$ and $\lambda_2$) 
to two-body meson scattering amplitude in a single channel, 
one can set 
$\lambda_1 \sim \mathcal{O}(N^{-2})$ and $\lambda_2\sim \mathcal{O}(N^{-1})$ 
in order to have a tenable perturbative expansion \cite{Manohar:1998xv}. 
$\mu^2\sim \mathcal{O}(1)$ because of no flavor degeneracy in a single channel. 
As consequences, one finds $m_{\phi,\chi}^2 \sim \mathcal{O}(1)$ 
thus $\tau \sim \mathcal{O}(1)$. 

Now we are ready to see what happens in Large $N$ limit.
After taking $N \to \infty$ as keeping $m_\phi$ and $m_\chi$ finite, 
Eq.~(\ref{eq:nagv2}) can be solved by 
\beq
g(r) = 1,
\eeq
and the other equation (\ref{eq:nagv1}) becomes
\beq
f'' + \frac{f'}{r}-\frac{f}{r^2}
-\frac{m_{\chi }^2}{2 } f \left(f^2-1\right) = 0.
\eeq
This is the equation for $k=1$ Abelian global vortex.
Note that $m_\chi$ is shown up in the equation and solutions are independent of 
$m_\phi$ unlike the case of Abelian global vortex given in Eq.~(\ref{eq:abelianvortex}).
The solutions themselves happen to be identical to those for $m_\phi = m_\chi$ ($\tau=1$) with
any finite $N$.

%%%%%%%%%%%%%%%%%%%%%%%%%%%%%%%%%%

\end{document}